\documentstyle[prb,aps,epsf]{revtex}
\begin{document}
\draft
\twocolumn[\hsize\textwidth\columnwidth\hsize\csname 
@twocolumnfalse\endcsname
\title{
Calculation of Densities of States and Spectral Functions
by Chebyshev Recursion and Maximum Entropy}
\author{R. N. Silver, H. R\"{o}der}
\address{MS B262 Theoretical Division, Los Alamos National
Laboratory, Los Alamos, New Mexico 87545}
\date{\today}
\maketitle
\begin{abstract}

We present an efficient algorithm for calculating spectral properties
of large sparse Hamiltonian matrices such as densities of states and
spectral functions. The combination of Chebyshev recursion and maximum
entropy achieves high energy resolution without significant roundoff error,
machine precision or numerical instability limitations. If controlled
statistical or systematic errors are acceptable, cpu and memory
requirements scale linearly in the number of states. The inference of
spectral properties from moments is much better conditioned for
Chebyshev moments than for power moments.  We adapt concepts from the
kernel polynomial method, a linear Chebyshev approximation with
optimized Gibbs damping, to control the accuracy of Fourier integrals
of positive non-analytic functions. We compare the performance of
kernel polynomial and maximum entropy algorithms for an electronic
structure example.

\end{abstract} 
\phantom{.}
]

\section{Introduction} 

Many computational physics problems involve very
large sparse Hamiltonian matrices. If $N$ is the number of states,
finding all eigenvectors and eigenvalues requires cpu time scaling as
$N^3$ and memory scaling as $N^2$. For individual eigenstates the
preferred method is Lanczos diagonalization, which uses only
matrix-vector-multiply operations and requires cpu and memory scaling
as $N$. Densities of states and spectral functions for finite
dimensional Hamiltonians are sums of $\delta$-functions with positive
amplitudes. In the thermodynamic limit of relevance to condensed
matter physics, these can extrapolate to singular structures such as
isolated states, band edges and Van Hove singularities. New linear
scaling methods are needed for calculating such spectral properties
which involve many eigenstates, and for quantities derived from them
including thermodynamics, total energies for electronic structure and
forces for molecular dynamics. Limited energy resolution
and statistical accuracy are often acceptable provided uncertainties
can be quantified.

The {\it maximum entropy method} \cite{mead,carlsson,drabold1} is a
popular approach: Maximize the information theoretic relative entropy
of the spectrum subject to data constraints. The input data are
usually power moments. Maximum entropy spectra are strictly
positive. Maximum entropy spectra are the solution of a convex
non-linear optimization problem.  Maximum entropy always yields 
broadened representations of the true spectra. The resolution function is
non-uniform and unknown, with some parts of a spectrum converging more
rapidly than others as the number of moments increases. Occasionally,
maximum entropy yields spurious structure; for example, it can `ring'
in smooth regions of a spectrum near to a Van Hove singularity. 
Non-analytic features are better approximated at higher energy
resolution, which is achieved in maximum entropy by fitting more
moments.  However, as moment order increases, the calculation of power
moments is more sensitive to machine precision limits, and the
optimization problem is more ill-conditioned. Maximum entropy is
difficult to implement for more than about fifty power moments.

The {\it kernel polynomial method} \cite{silver1,silver2,wang} is much easier
to implement for high energy resolution applications. It is a linear
Chebyshev approximation to spectra using Chebyshev moment data. Abrupt
truncation of Chebyshev series results in the Gibbs phenomenon: a lack
of uniform convergence at non-analytic (or singular) features in
spectra. Instead, the moments of the kernel polynomial approximation
are the data multiplied by Gibbs damping factors, which are chosen to
ensure positive spectra with the highest energy resolution. A kernel
polynomial approximation is a convolution of the true spectrum with a
known positive kernel polynomial function.  It can be rapidly
evaluated by fast Fourier transform without non-linear
optimization. In contrast to Lanczos methods, Chebyshev recursion is
numerically stable without accumulation of roundoff error; thus, there
is no need for computationally expensive reorthogonalization
\cite{parlett}. For sparse Hamiltonians, the computational cost for generating
Chebyshev moment data is linear scaling if controlled systematic or
statistical errors are acceptable \cite{silver1,voter}.  Chebyshev
approximations have been applied recently to densities of states and
spectral functions in diverse areas of condensed matter physics
including the Heisenberg antiferromagnet \cite{silver1}, the Holstein
$t-J$ model \cite{roder1}, the dielectric constants of quantum dots
\cite{wang2}, linear scaling algorithms for tight binding molecular
dynamics \cite{goedecker,voter}, non-orthogonal electronic structure
\cite{roder2}, and so on. Chebyshev approximations have also 
been developed independently for scattering problems in quantum
chemistry \cite{kosloff1,huang1,zhu1,kouri1,parker1,kouri2}. 

A comparison of the maximum entropy and kernel polynomial methods
reveals advantages for each.  (A comparison of Lanczos methods with
kernel polynomial methods may be found in Silver, {\it et al.}
\cite{silver2}.) The maximum entropy method achieves 
significantly higher energy resolution,
requiring calculation of four to ten times fewer moments for typical
applications. However, the non-linear optimization problem can be
difficult to solve, the resolution is non-uniform and unknown, and
there is a risk of artifacts.  The kernel polynomial method has
significantly poorer energy resolution. However, non-linear
optimization is not needed, the resolution function is uniform and
known, and there is no risk of artifacts.

Our experience is that, in most cases, the computational cost of
generating moment data limits the
ability to do calculations. The practical necessity to use
computational resources in the most efficient way 
motivates our development of a new maximum entropy
algorithm based on Chebyshev moment data.  Chebyshev moments have
several advantages over power moments for a maximum entropy algorithm:
\begin{itemize}
\item Machine Precision Limitations - 
In a power moment
information in low digits past the decimal point is redundant with 
information in low order
moments. New information is contained in higher digits whose
cardinality increases with the order of the moment. Thus, machine precision
puts a limit on how many power moments are useful to calculate.
In contrast, there is no redundancy in moments constructed from
orthogonal polynomials \cite{wheeler} and machine precision is
not limiting.
\item Conditioning - The 
ill-posed inverse problem of infering a spectrum from a limited
number of Chebyshev moments
is much better conditioned than from the same number of power moments.
In particular,
the Hessian for maximum entropy
optimization has a much flatter eigenvalue spectrum for Chebyshev
moments than for power moments.
\item Computational Efficiency and Accuracy - A simple coordinate
transformation converts a
Chebyshev series to a
Fourier series, which enables use of 
fast Fourier transform methods.
\end{itemize}
In summary, the combination of Chebyshev recursion and
maximum entropy should provide an efficient stable algorithm
capable of reaching arbitrarily high energy resolution.  

There is an extensive literature
\cite{auyeng,skilling,turek} on convex non-linear optimization applied
to maximum entropy.  For our applications, we find the principal new
algorithmic difficulty to be control of the numerical accuracy of
Fourier integrals when the true spectra have singular (or
non-analytic) features such as $\delta$-functions. We adapt concepts
from the kernel polynomial method and the Shannon sampling theorem to
solve this numerical accuracy problem. The resulting algorithm has no
difficulty handling thousands of Chebyshev moments, if necessary.

In Section II we briefly review methods for generating Chebyshev moment data.
In Section III we describe the kernel polynomial method.  In Section IV
we present our maximum entropy algorithm.  In Section V we illustrate the
method using an electronic structure example, comparing the performance of the
maximum entropy and kernel polynomial methods.  
In Section VI we conclude.

\section{The Generation of Chebyshev Moment Data}

Consider a density of states as representative of the spectral
properties of interest. The first step
is to scale the Hamiltonian, ${\bf H}=a{\bf X}+b$ such that
all eigenvalues $x_n$ of ${\bf X}$ satisfy $-1 \le x_n \le +1$. 
These endpoints are rapidly computed, for example, by Lanczos
methods using the same matrix-vector-multiply operations required
for generating Chebyshev moments. The only difference between
the kernel polynomial method and the maximum entropy method is,
in order to minimize endpoint corrections in fast
Fourier transform evaluation of Fourier integrals, we recommend
placing all $x_n$ well inside $-1$ and $+1$, for example
$-.99 \le x_n \le +.99$. This point will be discussed further
in Section IV.

The density of states is then
\begin{equation}
D(x)=\frac{1}{N}\sum_{n=1}^N \delta(x-x_n) \;\;\;.\label{q1}
\end{equation}
Data about $D(x)$ consist of 
Chebyshev moments,
\begin{equation}
\widehat{\mu}_m=Tr\{T_m({\bf X})\}=\int_{-1}^{1}T_m(x)D(x)dx \;\;\;.
\end{equation}
We use the notation $\widehat{\mu}_m$ for a datum on a moment, even
if this estimate is approximate. We calculate
moments using the Chebyshev recursion relation,
\begin{equation}
T_{m+1}({\bf X})=2{\bf X}T_m({\bf X})-T_{m-1}({\bf X})\;\;\;,
\end{equation}
which requires the same optimized matrix-vector-multiply algorithm
used in Lanczos methods.  Unlike Lanczos recursion, Chebyshev
recursions are numerically stable to arbitrarily large numbers of
recursions.  We use rules for multiplying Chebyshev polynomials,
e.g. $T_{2m}=2T_{m}T_m-1$, so that only $M/2$ matrix-vector-multiplies
are needed to generate $M$ moments.

Exact evaluation of $M$ moments requires cpu time proportional to
$O(N^2M/2)$ for sparse matrices. 
Generate $T_m({\bf X})|i>$ for each basis state
$|i>$. The estimator for Chebyshev moments is then
\begin{equation}
\widehat{\mu}_{2m}=\frac{2}{N}\sum_{i=1}^N <i|T_m({\bf X})T_m({\bf X})|i> - 1\;\;\;.
\end{equation}
There is a similar expression for $m$ odd.

Stochastic evaluation \cite{silver1} requires cpu time scaling as
$O(NMN_r)$. The estimator for Chebyshev moments is
\begin{equation}
\widehat{\mu}_m \approx \frac{1}{N_r}\sum_r <r| T_m({\bf X})|r>\;\;\;, 
\label{q216}
\end{equation}
where the $|r>$ are $N_r$ Gaussian random vectors. Such data have
statistical variance proportional to $(N N_r)^{-1}$ which may be
expressed directly in terms of the density of states.  
Estimation of statistical
errors is described in Silver and Roder \cite{silver1}.
More sophisticated choices of random vector 
appear to reduce statistical variance \cite{drabold1}, 
but they introduce unwanted statistical bias and make 
error estimation difficult.

Local truncation evaluation of moments requires cpu time scaling as
$O(N M J)$. Here, moments are calculated with a locally truncated
Hamiltonian, ${\bf H}_i$, where $J$ is the number of states included in
the truncation range. The estimator for Chebyshev moments is
\begin{equation}
\widehat{\mu}_m\approx \sum_i <i|T_m({\bf X}_i)|i>\;\;\;.
\end{equation}
This generates data with a systematic error determined by the
truncation range. `Logical' truncation \cite{voter} appears to
converge more rapidly and smoothly than `physical' truncation schemes
\cite{goedecker,goedecker2}. 
Local truncation may be applicable if the density matrix
has only local off-diagonal elements, as in tight-binding Hamiltonians
for insulators.  Exact moment derivatives needed to estimate forces
for molecular dynamics can be calculated from a Chebyshev derivative
formula.

Cpu time and memory limit the number of moments $M$ and their
statistical and systematic errors. Fortunately, both stochastic
and local truncation methods
provide means to estimate and control errors.

\section{The Kernel Polynomial Approximation}

The kernel polynomial method has two roles in this paper.  First, it
is a method to estimate spectra from Chebyshev moment data. Second, it
provides our approach to control numerical accuracy in the evaluation
of Fourier integrals of non-analytic spectra in our maximum entropy
algorithm.

An exact Chebyshev moment expansion of the density of states is
\begin{equation}
D(x)=\frac{1}{\pi \sqrt{1-x^2}}\left[\mu_0+2\sum_{m=1}^{\infty}\mu_m T_m(x)
\right]\;\;\;.\label{q8}
\end{equation}
The kernel polynomial method truncates Eq. (\ref{q8}) at $M$ moments,
introduces a factor $g_m^M$ to damp Gibbs phenomenon, and 
substitutes (possibly inaccurate) data $\widehat{\mu}_m$ for the
moments. The kernel polynomial approximation to a density of states 
is then
\begin{equation}
D_K(x)=\frac{1}{\pi \sqrt{1-x^2}}\left[1+2\sum_{m=1}^{M}\widehat{\mu}_m g_m^M 
T_m(x) \right]\;\;\;.
\end{equation}
Let $\phi\equiv\cos^{-1}(x)$. Then $T_m(x)=\cos(m\phi)$. Define $D(\phi)
\equiv \sin(\phi)D(X)$. The Chebyshev moments are then Fourier integrals,
\begin{equation}
\mu_m=\int_{-1}^{1} T_m(x) D(x) dx = 
\int_0^{\pi} \cos(m\phi) D(\phi) d\phi \;\;\;. 
\label{q717}
\end{equation}
If the data are exact,
$D_K(\phi)$ can be represented as both 
a simple convolution and a truncated Fourier series,
\begin{eqnarray}
D_K(\phi)=\int_0^{2\pi} \delta_K(\phi-\phi_o)D(\phi_o) 
d\phi_o\;\;\;,\nonumber\\
\delta_K(\phi)=\frac{1}{2\pi}\left[g_0+2\sum_{m=1}^M g_m^M \cos(m \phi )
\right]\;\;\;.
\label{q701}
\end{eqnarray}
For later purposes, we emphasize that the
Fourier integrals of $D_K(\phi)$ are $\mu_m=\widehat{\mu}_m g_m^M$. Thus,
$D_K(\phi)$ does not fit the moment data. 
If the data are inexact,
corresponding random variables should be added to Eq. (\ref{q701}).

The {\it kernel} $\delta_K(\phi)$ is a $2\pi$-periodic polynomial
approximation to a Dirac delta function, analogous to the resolution
function of a spectrometer. Resolution is uniform in $\phi$ with width
$\Delta \phi \propto M^{-1}$.  If $g_m^M=1$, at large $|\phi|$ the
kernel is oscillatory with period $\Delta \phi = \pi/M$ within an
envelope function decreasing slowly as $1/\phi^2$.  The result is the
Gibbs phenomenon: a lack of uniform pointwise 
convergence of the cosine series at singular (or non-analytic) structures
in the density of states.  
An optimal $g_m^M$ that minimizes the Gibbs phenomenon may be
derived \cite{silver2} by requiring the kernel to be a strictly positive
normalized polynomial of degree $M$ with minimal variance in $\phi$.
The result is
\begin{equation}
g_m^M=\sum_{\nu=0}^{M-m} a_{\nu}a_{\nu+m}\;\;\;,\label{q9}
\end{equation}
where
\begin{equation}
a_{\nu}=\frac{U_{\nu}(\lambda)}{\sqrt{\sum_{\nu=0}^M U_{\nu}^2
(\lambda)}}\;\;\;,
\end{equation}
and 
\begin{equation}
U_{\nu}(\lambda) 
=\frac{\sin((\nu+1)\phi_{\lambda})}{\sin(\phi_{\lambda})}
\;\;\;;\;\;\;\cos(\phi_{\lambda})=\lambda\;\;\;.
\end{equation}
Here $U_{\nu}$ are Chebyshev polynomials of the second kind
and $\phi_{\lambda}\equiv\pi/(M+2)$  The
$g_m^M$ decrease smoothly and monotonically 
from $1$ to $0$ as $m$ increases from $0$ to
$M$. This kernel was originally derived by 
minimizing the uniform norm \cite{jackson}.
Its envelope function decreases exponentially at large $|\phi|$.

The kernel polynomial method is also applicable 
to spectral
functions \cite{silver2},
\begin{equation}
A(\omega)=\lim_{\eta \rightarrow 0+} 
\frac{1}{\pi} Im \left\{ <\Psi_0|{\bf O}^{\dagger} 
\frac{1}{\omega - {\bf H}-i\eta} {\bf O}|\Psi_0>\right\} \;\;\;,
\end{equation}
where ${\bf O}$ is the appropriate Hermitian operator.
The corresponding Chebyshev moments have the form
$\widehat{\mu}_m^{\bf O}=<\Psi_0|{\bf O}^{\dagger}T_m({\bf X}){\bf O}|\Psi_0>$.
Silver {\it et al.} \cite{silver2} compare the 
performance of the kernel polynomial method to Lanczos methods
for spectral functions. 

Applications of kernel polynomial approximations 
to thermodynamics use a rapidly converging Fourier-Bessel
expansion of the partition function \cite{silver1},
\begin{equation}
Z=e^{-\beta b}\left[I_0(\beta a)+2\sum_{m=1}^{\infty} I_m(\beta a) \widehat{\mu}_m 
\right]\;\;\;.
\end{equation}
The $I_m(\beta a)$ are modified Bessel functions.
The partition function involves
integral rather than pointwise convergence, so
the optimal choice is no Gibbs damping, $g_m^M=1$.

Our maximum entropy algorithm uses the kernel polynomial approximation
in a novel way. We employ fast Fourier transforms to evaluate Fourier
integrals of $D(\phi)$ by summing over $L+1$ pixels equally spaced in
$\phi$, corresponding to sampling $D(\phi)$ at the $L+1$ zeroes of
$\cos((L+1)\phi)$. The Shannon sampling theorem says that the only
function which can be exactly evaluated by this procedure is a band
limited function, a finite $L$'th order cosine series. So, in fact,
our algorithm exactly evaluates Fourier integrals of a Chebyshev
approximation. The procedures in Section II generate moments of a
function consisting of a sum of $N$ $\delta$-functions with positive
amplitudes, equivalent to an infinite order Chebyshev series.
Inasmuch as, typically, $N \gg L$ we should not expect to resolve all
states. Our maximum entropy algorithm requires moments of an $L$-th
order positive Chebyshev approximation. The only Chebyshev
approximation that can satisfy the positivity constraint required by
maximum entropy is a kernel polynomial approximation. Moments of the
kernel polynomial approximation are related to moments of the true
function by Gibbs damping factors. This subtle difference becomes
important because we demand high numerical accuracy.  We elaborate on
these points in the next section.

\section{The Maximum Entropy Algorithm}

This section presents our maximum entropy algorithm.  Although
it may be regarded as an adaptation of previous maximum entropy algorithms
\cite{auyeng,skilling,turek}, our problem has novel issues
of numerical accuracy for $D(x)$ that contain singular 
(non-analytic) structures such
as $\delta$-functions.

Consider the case where the data are subject to Gaussian uncertainties,
\begin{equation}
\widehat{\mu}_m=\mu_m+\eta_m\;\;\;;\;\;\;{\bf E}\eta_m=0\;\;\;;\;\;\;
{\bf E}\eta_m\eta_{m'}=\sigma_m^2\delta_{mm'}\;\;\;.
\end{equation}
Here $\eta$ is a random variable and
${\bf E}$ denotes the statistical expectation value of the random
variable following it.
The $\chi^2$
statistic for measuring quality of fit is then
\begin{equation}
\chi^2=\sum_{m=0}^{M}
\left(\frac{\widehat{\mu}_m-\mu_m}{\sigma_m}\right)^2\;\;\;.
\end{equation}
In case of exact moment data, set $\sigma_m$ to the
numerical precision required, which can be very small. 
The $m=0$ term is included to constrain normalization,
$\widehat{\mu}_0=1$.  Taking the limit $\sigma_0 \rightarrow
0$ strictly enforces normalization.
In our applications using 32 bit computers, 
$6$th or $7$th digits past the decimal point 
of moments often contain important information.
We typically drop $\chi^2$ by
12 to 14 orders of magnitude below its starting values during the
course of converging to a maximum entropy solution.
Such high numerical accuracy can be critical to avoid
spurious artifacts and to yield 
the correct physics.

Therefore, very careful attention to numerical accuracy is
required in evaluating 
Fourier integrals, Eq. (\ref{q717}).
To have an efficient maximum entropy algorithm, we evaluate 
Fourier integrals by
fast Fourier transform. This equals
a sum over equally spaced points in $\phi$,
\begin{equation}
\mu_m \approx\sum_{l=0}^L\cos(m\phi_l)D(\phi_l)\Delta
\phi \;\;\;. \label{q211}
\end{equation} 
The $L+1$ $\phi_l$ satisfy
$\cos((L+1)\phi_l)=0$ where $0 \le l \le L$.
The Shannon sampling theorem says this approximation becomes
exact only if $D(\phi)$ is a band limited function of degree $L$. 
But the exact $D(\phi)$ in our applications are sums of $\delta$-functions
with positive amplitudes, so 
evaluation of Fourier integrals
by fast Fourier transform 
with a finite number of pixels $L$ is not exact. The maximum entropy
$D(\phi)$ also correspond to infinite order Fourier series, so evaluation
of their Fourier integrals by this procedure is not exact either.

Our strategy to minimize numerical errors is to minimize high
frequency components of the maximum entropy solution.  The goal of our
algorithm is to find a kernel polynomial approximation for $M \times
K$ moments instead of $M$ moments, where $K$ is some integer. Maximum
entropy provides the criterion for extrapolating the moment series.
But the moments of the kernel polynomial approximation of degree $M
\times K$ are $\widehat{\mu}_m g_m^{M \times K}$, so these are what we
should use as data in our $\chi^2$ criterion. Modifying the data in
this way ensures that our target spectrum is positive and satisfies
the Hausdorff conditions for the existence of a maximum entropy
solution.  Choosing $K$ in our algorithm is equivalent to choosing the
desired energy resolution.  If our maximum entropy solution was in
fact equivalent to a higher resolution kernel polynomial
approximation, choosing the number of pixels $L=M \times K$ would
yield exact Fourier integrals. But inasmuch as our maximum entropy
solution is not exactly band limited, we choose the number of pixels
some integer factor $I$ larger than $M \times K$, i.e. $L=M \times K
\times I$.  The extra factor $I$ further reduces numerical errors.

Increasing $L$ to improve numerical accuracy must be balanced against
increased computational resources required for the fast Fourier
transform. Cpu time scales as $O(L \ln L)$ and memory scales as
$O(L)$. Typically, we find maximum entropy improves resolution by
factors of 4 to 10 over the kernel polynomial method, so most of the
gain is obtained by choosing $4 \le K \le 10$.  The corresponding
$g_m^{M \times K}$'s for $0 \le m \le M$ are only slightly smaller than one,
but this difference is enough to determine whether our algorithm
converges to the stopping criterion or stalls at high $\chi^2$. Without
the Gibbs damping correction, convergence may be very non-uniform and
in some regions 
approach an energy resolution 
that can not be described with the number of pixels chosen. 
WIth the Gibbs damping correction, the dynamic range of the
resolution improvemnt is limited and can be handled with
the number of pixels chosen. We
also typically find the choice $I \ge 4$ to be sufficient to fit
Chebyshev moments to $7$ digits accuracy.

Endpoint corrections are another concern in evaluating Fourier
integrals.  They are often essential to get reasonable
convergence for high order moments. Sophisticated approaches to this
problem have been developed based on interpolation schemes
\cite{press}. However, within our algorithm we have the option to
minimize endpoint corrections at the outset, by forcing the spectrum
to be close to zero near endpoints $\phi=0$ and
$\phi=\pi$. The easy way to force $D(\phi)$ toward zero at endpoints
is to scale the Hamiltonian, ${\bf H}=a{\bf X}+b$ such that all
eigenvalues $x_n$ of ${\bf X}$ lie between, say $-.99$ and $+.99$,
rather than $-1$ and $+1$.  This change has only a $1$\% impact on
$\phi$ resolution, but it avoids the endpoint corrections and the bias of
interpolation schemes.  This justifies the recommendations made in
Section II for scaling the Hamiltonian. We often find this superficially small
correction to be critical to achieving high accuracy fits without
stalling.

Our optimization problem is to maximize the relative entropy,
\begin{equation}
S\equiv\int_{0}^{\pi}\left[D(\phi)-D_o(\phi) -
D(\phi)\ln\left(\frac{D(\phi)}{D_o(\phi)}\right)\right]d\phi
\;\;\;,
\end{equation}
subject to data constraints.  $S$ is strictly negative and equals zero
only when $D=D_o$.  Here $D_o(\phi)$ is a default model for the
density of states in the absence of data. We obtain faster convergence
with less risk of spurious artefacts by using prior knowledge to
choose a default model closer to the final answer. In the absence of
prior knowledge, we usually use the kernel polynomial approximation as
the default model, inasmuch as the motivation for maximum entropy is
to improve energy resolution beyond the kernel polynomial method.

More specifically, the {\it primal optimization problem} is to
maximize
\begin{equation}
Q_p \equiv S-\frac{\chi^2}{2\alpha}\;\;\;
\end{equation}
as a function of a continuous variable $D(\phi)$.
The statistical regularization parameter $\alpha$ sets a balance
between the fit, measured by $\chi^2$, and an information
measure, $-S$, of distance between the inferred $D(\phi)$ and the
default model $D_o(\phi)$. Alternatively, 
we regard $1/\alpha$ as a Lagrange
multiplier enforcing a constraint on $\chi^2$. 

Our algorithm consists of three nested loops: the outer loop
iteratively decreases $\alpha$ starting from a high value, until a
stopping criterion is reached; the middle loop for each $\alpha$
iteratively solves for the $D(\phi)$ that maximizes $Q_p$; the inner
loop at each $\alpha$ and $D(\phi)$ solves for the next update of
$D(\phi)$ using linear equation solvers such as conjugate gradients. 
We discuss each of these
loops in turn. 

The outer loop typically
starts at large $\alpha_1 \approx \chi_o^2$, the $\chi^2$ of the 
default model $D_o$. Then progress geometrically down in $\alpha$, e.g.
$\alpha_{k+1}=\alpha_k/2$. The corresponding
$\chi^2$ decreases and the information, $-S$, increases.
If the middle loop is unstable, as measured
by a significant increase in $\chi^2$, go back to conditions
at the start of this loop, halve the
step down in $\alpha$, and iterate until stability is reached. 
Popular stopping criteria for $\alpha$ are $\chi^2=M$ and
$\chi^2-2\alpha S=M$, although many other criteria are discussed in
the literature. 
Once the stopping criterion
is passed, perform a golden search for the optimal $\alpha$.
We often find that the information $-S$ saturates at an
$\alpha$-independent value as $\alpha$ decreases, so that the
outer loop may be stopped earlier.

In principal, the middle loop solves the primal optimization
problem,
\begin{equation}
\frac{\delta Q_p}{\delta D(\phi)}=-\ln\left(\frac{D(\phi)}{D_o(\phi)}
\right)
+\sum_{m=0}^M\frac{\widehat{\mu}_m-\mu_m}
{\alpha \sigma_m^2}\cos(m\phi)=0\;\;\;.\label{q6}
\end{equation}
This is a convex optimization problem having a unique answer.
Unfortunately, this approach is difficult because this
problem statement is written in terms of a continuous positive
variable $D(\phi)$ which varies typically by many orders of magnitude.
In practice, the number of variables to optimize would be the number of 
pixels $L$ chosen, which for reasons of numerical accuracy we discussed
earlier is usually a quite large number.
However, a {\it dual optimization problem} \cite{auyeng} 
solves the same problem, is
more stable numerically, and is easier to implement
than the primal problem. It requires only $M \ll
L$ parameters $\vec{\lambda}$ defined by
\begin{equation}
\lambda_m\equiv\frac{\mu_m-\widehat{\mu}_m}{\alpha \sigma_m^2}
\;\;\;.\label{q3}
\end{equation}
Then the maximum entropy $D(\phi)$ satisfying Eq. (\ref{q6}) is
\begin{equation}
D(\phi)=D_o(\phi)\exp\left(-\sum_{m=0}^M \lambda_m \cos(m\phi)\right)
\;\;\;.\label{q214}
\end{equation}
This form is also obtained by maximizing entropy subject to
Lagrange contraints on moments with Lagrange multipliers
$\vec{\lambda}$. 
The dual optimization problem is to maximize
\begin{equation}
Q_d\equiv\ln\left(\int_0^{\pi}D(\phi)d\phi\right)+
\sum_{m=0}^M \left[ \widehat{\mu}_m\lambda_m+
\frac{\alpha\sigma_m^2\lambda_m^2}
{2}\right]\;\;\;\label{q4}
\end{equation}
as a function of the set of $\lambda_m$. Mapping onto the dual space of 
Lagrange multipliers reduces
an infinite dimensional optimization problem to a feasible
finite dimensional problem. 
Away from the maximum, define 
\begin{equation}
\xi_m \equiv \frac{\partial Q_d}{\partial \lambda_m} =
\widehat{\mu}_m-\mu_m+\alpha\sigma_m^2\lambda_m\;\;\;.
\end{equation}
Eq. (\ref{q3}) is satisfied when $\xi_m=0$.

The middle loop of our algorithm solves
Eq. (\ref{q3}) by Newton-Raphson iteration. Beginning
with some starting $\vec{\lambda}^0$, the $n+1$'st step is
\begin{equation}
{\bf {\cal H}}_n(\vec{\lambda}^{n+1}-\vec{\lambda}^n)= \vec{\xi}^n
\;\;\;.\label{q213}
\end{equation}
Here ${\bf {\cal H}}$ is the Hessian of the dual problem, which
is a positive definite $M \times M$
matrix and a simple function of the moments,
\begin{equation}
{\cal H}
_{mm'}\equiv \frac{\partial^2 Q_d}{\partial \lambda_m \partial \lambda_m'}
=\frac{\mu_{m+m'}+\mu_{|m-m'|}}{2}
+\alpha\sigma_m^2\delta_{mm'}\;\;\;.\label{q215}
\end{equation}
Then, 
\begin{equation}
Q_d=Q_p+\sum_{m=0}^M \frac{\xi_m^2}{2\alpha\sigma_m^2}\;\;\;.\label{q5}
\end{equation}
We have $Q_d > Q_p$ in
Eq. (\ref{q5}), and as the iteration proceeds $Q_d$ ($Q_p$) approach
$Q_{\infty}$ from above (below).
Hence converging bounds at the $n$'th iteration are
$Q_d^n \ge Q^{\infty}\ge Q_p^n$ where
$Q^{\infty}\equiv \lim_{n \rightarrow \infty} \{Q_d^n,Q_p^n\}$.
These bounds provide stopping criteria for the middle loop. We typically
stop at $Q_d = 1.02 Q_p$.

The inner loop 
solution of Eq. (\ref{q213})
by linear equation solvers such as conjugate gradients.
This task 
can be handled by standard packages such as EISPACK. An advantage of
Chebyshev moments is that
the spectrum of eigenvalues of the Hessian in Eq. (\ref{q215}) is almost flat,
whereas the eigenvalue spectrum for power moments is steeply decreasing. 
Thus, this problem is well-conditioned for Chebyshev moments but
can easily become ill-conditioned for power moments as $M$ increases.

We find the cpu time required by our algorithm scales as
$O(ML\ln(L))$. But it remains negligible compared to the cpu time
required to generate moment data for most problems. As we stated
before, use of the maximum entropy method usually cuts overall cpu
requirements by at least a factor of $4$ over the kernel polynomial
method.  Isolated features in spectra, such as individual states and band
edges, usually converge much faster, up to a factor of $10$ or more.

A few words should be said about data generated by stochastic
methods in Eq. (\ref{q216}). Calculation of the covariance 
matrix ${\bf C}$ for such data is described
in an earlier paper \cite{silver1}. Its
structure is the same as the Hessian in Eq. (\ref{q215}). 
The appropriate generalized $\chi^2$ statistic is

\begin{equation}
\chi^2=( \widehat{\mu}-\mu)^{\dagger}{\bf C}^{-1}
( \widehat{\mu}-\mu)\;\;\;.
\end{equation}
There is a cancellation, leading to an effective Hessian 
essentially proportional to a unit matrix
and independent of $D$. This property
further facilitates finding maximum entropy solutions 
for data generated by the
stochastic method.

Energy derivatives needed for molecular dynamics can
be derived for maximum entropy using the same expressions for
exact derivatives of moments. The statistical error for stochastic methods
using Gaussian random vectors can easily be accomodated, because the
covariance of moments is proportional to the Hessian. Details
of these extensions will be presented elsewhere.

\begin{figure*}[t]
\includegraphics{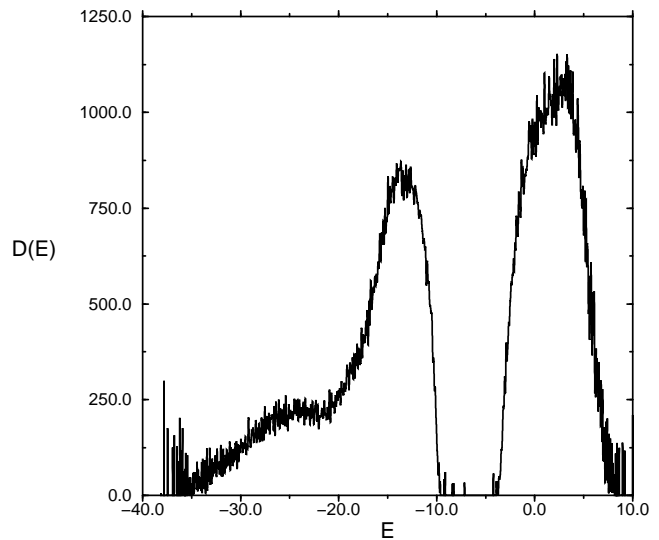}
\vbox to 3.3in{}
\caption{Density of states of an orthogonal tight binding model for
amorphous diamond calculated with the maximum entropy method using
$1024$ exact Chebyshev moments.}
\end{figure*}

\section{An Example}

We illustrate the performance of our MEM algorithm with the example of
an orthogonal tight binding calculation of the density of states of
amorphous diamond by Dong and Drabold \cite{dd}. The purpose of their
study was to examine the localized to extended transition in band-tail
states in an amorphous semiconductor. We refer readers to their paper
for a discussion of the physics. Here, we are only interested in a
comparison of methods for estimating densities of states. We note that
their maximum entropy calculations used Chebyshev moments but were
limited to about $90$ moments because of the difficulties in
implementing a maximum entropy algorithm. 

\begin{figure*}[t]
\includegraphics{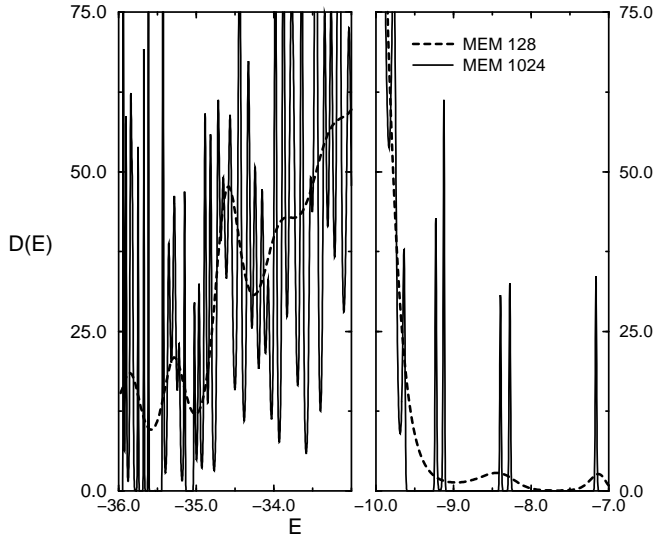}
\vbox to 3.3in{}
\caption{Comparison of portions of the density 
of states in Fig. 1 calculated by
maximum entropy for $128$ and $1024$ moments.}
\end{figure*}

\begin{figure*}[t]
\includegraphics{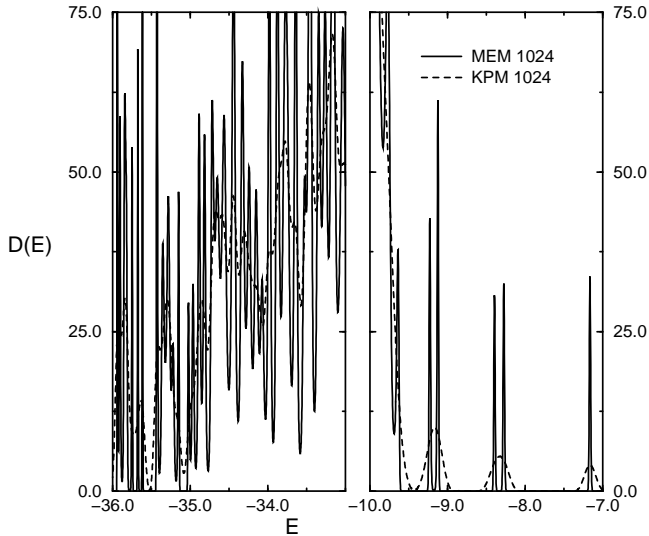}
\vbox to 3.3in{}
\caption{Comparisons of portions of the density of states in Fig. 1 calculated
by the kernel polynomial method (KPM) for $1024$ moments and by 
the maximum entropy method (MEM) for $1024$ moments.}
\end{figure*}

With the algorithmic
improvements presented in this paper, we demonstrate that maximum
entropy calculations with thousands of moments are now feasible.

The Hamiltonian considered has $16384$ states. We show maximum entropy
results for 
the goal of improving resolution by a factor of $8$ over the kernel
polynomial method. Hence, we choose
$K=8$. We set $I=4$ to minimize numerical errors in the
evaluation of Fourier integrals. We use up to $M=1024$ exact Chebyshev
moments. Hence the number of pixels $L = M \times K \times I \le 32768$. 

Fig 1 shows the full density of states obtained using 
maximum entropy for $M=1024$.
Note the optical gap in the density of states. Fig. 2 shows a
comparison of maximum entropy results for $M=128$ and 
$M=1024$ for two portions 
of this spectrum. On the left is a region where the states are dense,
and on the right is a region inside the gap where the states are
sparse. $M=128$ is a factor of two or three larger than 
the maximum number of moments 
that can be handled by maximum entropy algorithms using power moments. Our
algorithm clearly is able to handle many more moments and to achieve
much higher energy resolution. Figure 3 shows the same comparison
between kernel polynomial method and 
maximum entropy with $M=1024$
moments. Again maximum entropy provides a
dramatic improvement in energy resolution
over the kernel polynomial method.

\section{Conclusion} 

We have described an efficient algorithm to calculate densities of states and
spectral functions of large sparse Hamiltonians using Chebyshev
recursion and maximum entropy. It is a non-linear extension of the
kernel polynomial method. It is capable of handling large numbers of
moments and non-analytic (singular) structures in densities of states
and spectral functions to achieve high energy resolution. The choice
of Chebyshev recursion overcomes problems of machine precision and
ill-conditioning found in maximum entropy algorithms for power
moments. It also circumvents the accumulation of numerical roundoff
errors in Lanczos recursion. It controls the numerical accuracy of
Fourier integrals by multiplying the moment data by Gibbs damping
factors appropriate to the number of pixels chosen. It avoids endpoint
corrections to the fast Fourier transforms by appropriate scaling of
the Hamiltonian. Our algorithm achieves significant resolution gains
over the kernel polynomial method for practical physics examples. The
cpu time we need to find the maximum entropy solution scales
approximately as the number of pixels times the number of moments.
For most applications, this time is small compared to the cpu time we
need to generate the Chebyshev moment data. Overall cpu time can scale
linearly in the number of states if controlled statistical or
systematic errors are acceptable.

The resulting maximum entropy algorithm has some tuning parameters to
control the balance between numerical accuracy, convergence, energy
resolution and computational resources. A fortran 77 program
implementing our algorithms for KPM and MEM is available from the
authors by sending e-mail to rns@loke.lanl.gov. It uses publically
available libraries including DFFTPACK for fast Fourier transforms and
EISPACK for solving systems of linear equations.

\subsection*{\bf Acknowledgements}
Research supported by the U. S. Department of Energy. We thank D.
Drabold, J. J. Dong for kind permission to use the example
reported here. We thank D. Drabold and J. Kress for helpful comments
on the manuscript.

\end{document}